\documentclass[12pt]{article}
\usepackage{capt-of}
\usepackage{caption}
\usepackage{graphicx}
\usepackage{amsmath}
\usepackage{amssymb}
\usepackage{amsthm}
\usepackage{latexsym}
\usepackage{color}
\usepackage{graphicx}
\usepackage{color}
\usepackage{multirow}
\usepackage{eurosym}
\usepackage{subcaption}

\DeclareSymbolFont{calletters}{OMS}{cmsy}{m}{n}
\DeclareSymbolFontAlphabet{\mathcal}{calletters}

\def\be{\begin{eqnarray}}
\def\ee{\end{eqnarray}}

\def\b*{\begin{eqnarray*}}
\def\e*{\end{eqnarray*}}

\makeatletter \@addtoreset{equation}{section}

\newcommand{\un}{1\hspace{-1mm}{\rm I}}   

\usepackage{color}

\def \E{\mathbb{E}}
\def \F{\mathbb{F}}

\def \P{\mathbb{P}}

\def \R{\mathbb{R}}

\def\Fc{{\cal F}}

\def \0{\mathbf{0}}

\newcommand{\esp}[1]{\mathop{\rm \mathbb{E}\left (\mathit{#1} \right )}}

\def\1{{\bf 1}}

\title{The  Asset Liability Management problem of a nuclear operator : a numerical stochastic optimization approach. }

\author{ Xavier Warin \thanks{EDF R\&D \& FiME, Laboratoire de Finance des March\'es de l'Energie, xavier.warin@edf.fr}}

\date{\today}

\begin{document}

\maketitle

\abstract{
	We numerically study an Asset Liability Management problem linked to the decommissioning of French nuclear power plants. 
        We link the risk aversion of practitioners  to  an optimization problem. 
        Using different price models we show that  the  optimal solution is linked to a de-risking management strategy similar to a concave strategy and we
        propose an effective heuristic to simulate the underlying optimal strategy.
        Besides we show that the strategy is stable with respect to the main parameters involved in the liability problem.
}

\vspace{5mm}

\section{Introduction}
\subsubsection*{Strategies with constant weights}
        The  goal of long term Asset Management is to find an optimal allocation strategy  in some financial risky assets. 
        Asset Management has generated a lot of publications since the early work of Markowitz \cite{Markowitz} leading to the development of
        modern portfolio theory. In this framework, the risk management is achieved by an  optimal portfolio allocation in term of mean-variance. 
        This allocation is known to be very sensitive to inputs data \cite{Michaud} and
        the portfolio  may behave badly if the assets in the portfolio deviates from the estimated behavior. In order to deal with this robustness problem, some other allocation strategies  have been developed. Among them, equi-weighted portfolio (same weight in term of the amount invested in the different assets in the portfolio)  are often more effective on empirical  data  than
         any other strategy tested in \cite{DeMiguel} including methods with Bayesian approach \cite{James},\cite{Jorion}, \cite{Pastor} and variance minimization methods.
         In fact it is shown in \cite{Platen}, that this equi-weighted portfolio is a proxy for the Growth Optimal Portfolio  first studied in \cite{Maclean},\cite{Latane}, \cite{Breiman}, \cite{Thorp}, \cite{Markovitz1}, and later in \cite{long}, \cite{Cover}, \cite{Ziemba}, \cite{Brown}, \cite{Stutzer} \cite{Maclean}.
       Another approach is the risk parity allocation where a given level of risk is shared equally between all assets in the portfolio \cite{Roncalli}.
       Apart this risk allocation approach some investment strategies have been  developed and can be classified as explained below.
 \subsubsection*{Concave and convex strategies}
        Following \cite{Bruder}, the performance of a portfolio $V_t$
        between $t=0$ and $t=T$  invested in a risky asset 
       $S_t$  following a Black Scholes model \cite{Black} with trend $\mu$ and volatility  $\sigma$ and a non risky asset $B_t$   with a proportion $\pi$ invested in the risky asset
      can be decomposed as $\frac{V_T}{V_0} = A B C$ where: 
       \begin{itemize}
       \item  $A = e^{\Pi(S_T)-\Pi(S_0)}$ with $\Pi(x) -  \Pi(y) = \int_{y}^x \frac{\pi(s)}{s} ds$  an optional profile only depending on the strategy $\pi$ and the initial and final values of the asset $S_0$ and $S_T$,
       \item  $B= e^{-r \int_0^T [1- \pi(S_t)]dt}$ is the gain due to the investment in the risk free asset,
       \item  $C= e^{\frac{\sigma^2}{2} \int_0^T [ \pi(S_t) -\pi(S_t)^2 - S_t \pi'(S_t) ]dt}$ is the trading impact on the portfolio.
       \end{itemize}
       As explained in \cite{Bruder},  when the optional profile is convex, the volatility has a negative impact on the performance of the portfolio, and qualitatively the strategy consists in buying some risky assets when the price are increasing and selling them when the risky price is decreasing.
       When the optional profile is concave, the strategy often corresponds  to buying when prices are decreasing and selling when prices are increasing.\\
       This separation between convex and concave optional profile can often be used to identify the sell/buy behavior corresponding to a strategy.
       Between the most commonly used strategies we can  identify :
       \begin{enumerate}
       \item The ones without capital protection, some classical ones being :
         \begin{itemize}
           \item the Constant Mix strategies where  $\pi \in [0,1]$ is  a constant function \cite{Merton}.
          \cite{Merton1} showed that this strategy is 
         optimal with a Constant Relative Risk Aversion utility function. It  can be easily checked that the optimal profile of such a strategy is concave with a positive trading impact.
         \item the mean reverting strategy, buying the asset when its value is below a value $\bar{S}$ and selling otherwise. 
         The optional profile of this strategy is concave, and  the trading impact is positive for $\phi \in [0,1]$.
         \item the average down strategy choosing an investment strategy $\pi(V_t) = \alpha \frac{ \bar{V}- V_t}{V_t}$ where $\bar{V}$ is the portfolio target at date $T$ and 
$\alpha >0$. When the target is reached, the whole portfolio is invested in the risk free asset. Its optional profile is  concave with a positive trading impact. 
         \item the trend following strategies with the simple following principle : keep a risky asset as long as its trend  goes up and sell it when it goes down. \cite{Bruder} and \cite{potters} showed that
           this strategy could lead to a high  average return but can lead to high losses with a high probability,
         \item the regimes-bases portfolio management supposing that at least  two regimes   characterize  the asset behavior (see \cite{Ang} for the case with two regimes)  :
           one regime with  low  expected returns, high volatilities and high correlations, and 
           one regime with higher expected returns,  and lower volatilities and correlations. \cite{Ang} \cite{Ang1} \cite{Guidolin} \cite{Tu} showed that taking into account regimes can have
           a significant impact on  portfolio management. According to \cite{Ang}\cite{Ang1}, this impact is negligible if risk free assets are unavailable but becomes important otherwise.
           \cite{Guidolin} \cite{Tu} showed that taking into account different regimes provide higher expected returns.
         \end{itemize}
       \item The second class of strategies are the ones with capital protection, such as :
         \begin{itemize}
         \item the buy and hold strategy where the portfolio is invested 
               initially with a proportion $\pi$ in the risky asset, thus the minimum value of the portfolio at date $T$ is given by the actualized value of the investment in the risk free  asset,
             \item the stop loss strategy based on  a threshold $\bar{S}$ and a given strategy $\pi(S_t)$ and using the strategy $ \bar{\pi}(S_t)= \pi(S_t) \un_{S_t \ge \bar{S}}$. If $\pi$ is a concave strategy, 
               it can be shown that the stop loss strategy deteriorates the trading impact. Besides, this strategy provides a payoff identical to the one of a call option but starts with a smaller initial cost. \cite{Carr} showed that this strategy is in fact not self financing.
             \item the CPPI (Constant Proportion Portfolio Insurance) strategy, theoretically studied in \cite{Perold} \cite{Perold1} \cite{Black} \cite{Black1}, which is a convex strategy selling low and buying high. 
This strategy relies on a bond floor $F_t$ which is the value below which the portfolio values  should never go  in order to be able to ensure the payment of all future cash flows. It also relies on a multiplier coefficient $m$  such that the amount invested in the risky asset is $m(V_t-F_t)$. 
Using a coefficient $m=1$ gives a Buy-and-hold strategy, whereas $m<1$, $F_t=0$ gives a constant mix strategy. In practice, due to non continuous re-balancing or jumps \cite{Cont},
 there is a risk gap \cite{Amenc}
 meaning that the portfolio value may fall under the floor. Besides, the risk gap is exacerbated for high values of   $m$  and the portfolio turnover is higher than for Constant Mix strategies for example.
             \item the OBPI strategy  developed in \cite{Leland} which consists in  choosing $\pi_0$, in investing $(1-\pi_0) V_0$ in the risk free asset and in investing $\pi_0 V_0$ in a call option with maturity $T$ and strike $K$. The strategy is static thus there is no trading impact and only the implied  volatility of the option intervene in its valorization.
               The expected return of this convex strategy is increasing with $K$ but this increase in performance is achieved at the cost of higher probability of a null return. In the Black Scholes framework, the OBPI strategy is optimal when using a CRRA utility function \cite{ElKaroui}. In practice, implied volatility of options are higher than empirical one, which decreases the expected return of the strategy \cite{Zagst}.
One difficulty of this approach comes from the fact that the options are only available for short maturities forcing to use some rolling OBPI with high re-balancing in the portfolio. Some comparison between CCPI are OBPI  have been conducted in \cite{Bertrand} \cite{Bertrand1} \cite{Zagst} without clearly showing that one outperforms the other.
         \end{itemize}

       \end{enumerate}
\subsubsection*{ALM strategies}
All the previous methods only deals with the problem of Asset Management without constraints except the one trying to be above a fix capital at a given date. Generally, Asset Liability Management deals
with  the management of a portfolio of assets under the condition of covering some future liabilities as in the context of pension plans or insurance. Typically we are interested in covering liabilities at dates between 0 and 100 years where inflation and the long term interest rates  affect the liability: 
this constraints prevent the manager to use a cash flow matching technique  consisting in using some assets trying to replicate the liability because of the scarcity of the products to hedge  interest rates and inflation risks.\\
The literature on Asset Liability Management is far poorer than the one on ``pure'' Asset Management. 
Most of the strategies are  Liability-driven Investment strategies :  they  focus on hedging the liability \cite{Martellini} instead of trying to outperform a benchmark. A LDI strategy
typically splits the portfolio in two parts : a first part is a liability-hedging portfolio, while the second part is a performance-seeking portfolio. The first portfolio has to be highly correlated to the liability
while the second one has to be  optimal within  the mean-variance approach for example.
In this framework, some adaptations of the previously described Asset Management methods with capital protection have been developed for the liability problem :
\begin{itemize}
\item the  CPPI strategy  has been adapted to give the Core-Satellite Investing strategies \cite{Amenc} where the risk free asset is replaced by a hedging portfolio and the risky asset replaced by  a efficient portfolio in the Markowitz approach.
\item The OBPI strategy can be adapted following the  same principle \cite{Martellini1}.
\end{itemize}
From the theoretical point of view, following Merton's work \cite{Merton1},  many ALM problems  have been treated 
but forgetting the regulatory constraints on the fund due to liabilities \cite{Sundaresan} \cite{Rudolf} \cite{Detemple}.
Most of the time  some funding ratio constraints are imposed by  regulation and not a lot of articles have taken them into account \cite{Martellini1},\cite{Hoevenaars}, \cite{Binsbergen}, \cite{Wang}.
\subsubsection*{The ALM problem of the French nuclear operator}
The problem we aim to solve is the ALM problem faced by nuclear power plant operators : in some countries, regulation impose to the operators to hold
 decommissioning funds
in order to cover the future cost of dismantling the plants,  and treating the nuclear waste.
In France the laws 2006-739 of June 28, 2006  and  2010-1488 of 7 December 2010  on the sustainable management of radioactive materials and waste impose to the French operators to hold such a fund.
In order to estimate the liability part,  the future cash flows are discounted with a discount rate complying with regulatory constraints and indexed by a long term rate (TEC 30) averaged on the past $10$ years.
Besides, this value must be coherent with the expected returns of the assets : ``the interest rate can not exceed the portfolio return as anticipated with a high degree of confidence''.\\
The fund has been endowed by payments until the beginning of 2012 giving in 2016   a funding ratio (portfolio value divided by liability value) of $105\%$ thus it is estimated to be sufficient to cover at this date the discounted value of the future cost \cite{coursDesComptes}. In this article we will suppose  that at the current date we are at equilibrium so the funding ratio is equal to $100\%$.\\
The liability constraint is imposed every 6 month.\\
The liability part is subject to some risks :
\begin{itemize}
\item the first (and  most important  for the first years) is the risk due to the long term rate : a shift in the long term rate can trigger a constraint violation causing a refunding obligation,
\item the second one is the inflation risk that can be important for very long term liabilities,
\item the last one is the uncertainty linked to the future charges linked to decommissioning costs.
\end{itemize}
The third risk is beyond the scope of this study, the second risk is of second order for reasonable models \footnote{A study using an inflation driven by a reasonable mean-reverting process has been conducted showing the small impact of the parameter at least for the first 10 years.}.
The first risk is hard to tackle numerically due to the non markovian dynamic of an average rate. Therefore we will deal with this risk by some
sensibility analysis.\\
As for the portfolio part, a pure hedging strategy cannot be used : in addition to the fact that inflation cannot be hedged in the long term with market instrument, a pure hedging strategy would not  give the necessary expected return to match the liability value.
In this article we suppose that 
a Constant Mix strategy is used for the portfolio : $50\%$ are invested in bonds while the other part is invested in an equity index.\\
On top of classical mean-variance measures, the risk measure often used by practitionners is the asset-liability deficit risk at a chosen confidence level.
As we will show, this risk aversion  will allow us to define some utility functions and an optimization problem associated.
Dealing with the investment problem with different models we will exhibit optimal strategies linked to this problem.\\
A simplified version of this problem has been recently adressed theoretically  in \cite{tankov} using  shortfall risk constraints.\\
The structure of the article is the following :
\begin{itemize}
\item We first describe the problem,
\item Then we suppose that the equity index  follows a Black Scholes model and we propose an objective function to model the risk aversion. 
We are able to give an heuristic to simulate the optimal strategy obtained and we show that this strategy is robust with respect to the long term discount factor used,
\item At last we suppose that the equity index follows an alternative model (MMM) and show that with the same objective function the strategy obtained is quite similar to the one obtained by the Black Scholes model.
\end{itemize}

\section{Describing the ALM problem }
\subsubsection*{The liability}
We model the problem in a continuous framework.
Supposing that the inflation $\gamma$ is constant, the discounted value $L_t$  of the liability by the risk free rate  at date $t$  can be written as :
\begin{equation}
L_t = e^{(\gamma-r)t} \sum_{t_j > t/ t_j \in \mathop{R}}  \hat D_{t_j} e^{-  a_L(tj-t)},
\end{equation}
where $r$ is the risk free rate supposed constant, $ a_L$ is the long term actualization factor that we take constant and $\hat D_{t_j}$ are
the values of the future payments at date $t_j$. The set $\mathop{R}$ defines the set of future dates of payment estimated for
decommissioning.\\
The payment dates are scheduled every month and most of the amount are scheduled to be paid  in more than 10 years and less than 20 years.
In table \ref{decom}, we give the estimated futures decommisionning charges as given in \cite{tankov}, \cite{coursDesComptes}.
\begin{table}[!h]
\centering
\begin{tabular}{|l|l|l|l|l|l|l|l|} \hline
Year & 2015 & 2020 & 2025 & 2030 & 2035 & 2040 & 2045 \\  \hline
Cash flow, M \euro & 200 & 950 & 5550 & 7950 &  2700 & 1500 & 500\\ \hline
\end{tabular}
\caption{\label{decom} Estimation of future decommissioning charges (5 years periods).}
\end{table}
\subsubsection*{The asset portfolio}
We note $A_t$  the actualized value of the portfolio by the risk free rate and we suppose that the portfolio is composed of assets invested
in bonds (so with a zero actualized return) and in an equity index with an actualized value $S_t$.\\
The liability constraint is imposed twice a year, defining the set $\mathcal{L}$ of the  dates where the portfolio value has to be above the liability value. We suppose that the constraint is imposed straightforwardly  at once without penalty.\\
In the sequel we will use two different models to model the equity index :
\begin{itemize}
\item the first one is the classical Black Scholes model \cite{Black} that is widely used but not adapted to long term studies,
\item the second model we will use is the MMM model developed by Platten in the Benchmark approach \cite{Platen} \cite{Platen1} which is  shown  to be more adapted to the long term modelization.
\end{itemize} 
\subsubsection*{The regulatory constraint}
Noting $D_t$ the discounted value of cumulated endowments until date $t$, we note 
$$P_t = A_t -D_t-L_t,$$
 the netted value of the portfolio.
We suppose that the endowment is only realized when the regulatory constraint is activated and that the endowment is realized in such a way that it is minimal at each date, so
the $D_t$ dynamic is given by :
\begin{equation}
dD_t = 1_{t \in \mathcal{L}} (L_t-A_t)^{+}
\end{equation}
\subsubsection*{The objective function}
From a practical point of view,  long term management of such a portfolio is made with a high risk aversion to large amounts of injections in the very long term. It corresponds
mathematically to an aversion to heavy tail in the left hand side of the distribution of $P_T$. 
These heavy tails in the negative values of the distribution of $P_T$ are symptomatic  of large endowments and typically  a linear minimization  of the losses  such as minimizing $\E(P_T)$ doesn't match the practitioners aversion.
This can easily explained by the cost of refinancing : heavy endowments can decrease the firm's rating causing an increase in the cost of new debt issuance.\\

So the objective function mostly penalize the negative tails in  the distribution of $P_T$ .
Using the two models allows us to test the sensitivity of the strategy obtained with respect to the modelling.
\subsubsection*{Main parameters of the study}
In the sequel, the $\gamma$ and $r$ parameters will be taken equal to $2\%$ annually, the classical value for the long term actualization is taken equal
to $2.6\%$ annually if not specified. The maturity of the study $T$ will be $20$ years so that $240$ dates of payment in $\mathop{R}$ are involved and $40$ dates
of constraints in $\mathcal{L}$ are imposed. The initial value of the fund is $23.35$ billions of Euros so that the initial funding ratio is equal to one.
\\
In the article, we shall consider a one dimensional Brownian motion $W$ on a probability space $(\Omega,\Fc,\P)$ endowed with the natural (completed and right-continuous) filtration $\F=(\Fc_t)_{t\le T}$ generated by    $W$ up to some fixed time horizon $T>0$.

\section{Optimal strategy with an equity index following  the Black Scholes model}
In this section, we suppose that the actualized index follows the Black Scholes model :
\begin{flalign*}
d S_t =  S_t ((\mu-r) dt + \sigma dW_t),
\end{flalign*}
where $\mu=7\%$ annually, the volatility $\sigma=18\%$  and  $W_t$ is a brownian motion.\\

\begin{figure}[h!]
\centering
\includegraphics[width=0.5\textwidth]{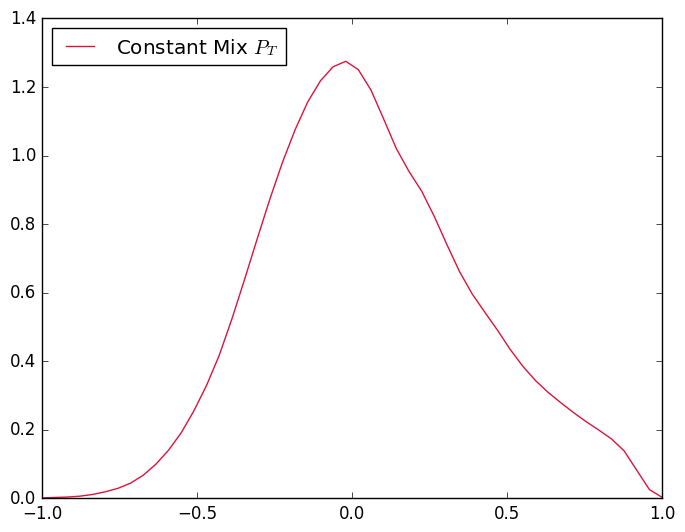}%
\captionof{figure}{\label{CMixBS} Normalized $P_T$ distribution for Constant Mix strategy  for Black Scholes model.}%
\end{figure}

Using the Constant Mix strategy, the normalized $P_T$ distribution such that the maximal loss is equal to $-1$ is  given on figure \ref{CMixBS}, so  one may  wonder how to  get a strategy  permitting to reduce the risk of high endowments.
In the sequel this normalization value will be used for all the distribution quantiles and figures.\\
We note $\phi= (\phi_t)$ the proportion of the portfolio $A_t$ invested in the equity index at date $t$. 
The portfolio dynamic is the given by :
\begin{eqnarray}
d A_t = \phi_t A_t ((\mu-r) dt + \sigma_t dW_t) + dD_t.
\end{eqnarray}
We propose to use an objective function :
\begin{equation}
J(t, A_t, D_t) = \esp{ g(-P_T) ~|~\Fc_t},
\label{eq:objective}
\end{equation}
where $g$ is function  with support on $\R^+$  and that we will suppose convex.
The optimization problem is then :
\begin{equation}
\hat J(t, A_t, D_t) =  \min_{\phi} \esp{ g(-P_T) ~|~\Fc_t}.
\label{Eq:OptimBS}
\end{equation}
In order to solve \eqref{Eq:OptimBS}, the deterministic Semi-Lagrangian methods \cite{warin} have been successfully used up to a maturity of 10 years.
An alternative used for the results presented here consists in discretizing $A_t$ and $D_t$ on a grid and calculating the expectation involved by Monte Carlo.
Using the Stochastic Optimization Library StOpt \cite{StOpt}, the calculation was realized on a cluster with MPI parallelization.
The grids for the asset discretization used a step of $200$ millions euros, while the endowment level is discretized with a step of $500$ millions Euros. A linear interpolation is used to interpolate a position in the bi-dimensional grid $(A,D)$. The number of simulations used to calculate expectations is chosen equal to $4000$.\\
The optimization part is followed by a simulation part using the optimal control calculated using $50000$ simulations.

\subsection*{Selection of the objective function to fit risk aversion}
On figure \ref{figFuncOptimBS} we give the normalized distributions  obtained by different objective functions $g(x)= g_1(x):=(x+ 0.4 x^2) 1_{x>0}$,
$g(x)=g_2(x):=(x+ 0.4 x^2+ 4e^{-4} x^3) 1_{x>0}$, $g(x)=g_3(x):=g(x) =(x+ 0.4 x^2+ 4e^{-5} x^3) 1_{x>0}$ :

\begin{figure}[h!]
\begin{minipage}[b]{0.5\linewidth}
  \centering
 \includegraphics[width=0.9\textwidth]{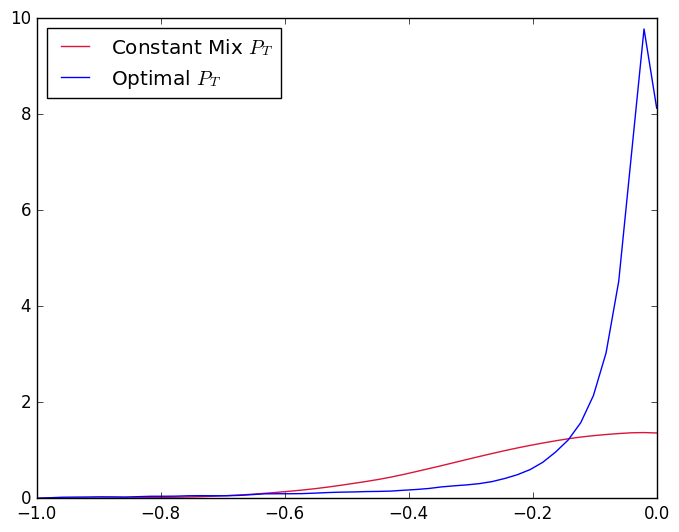}
 \caption*{$g_2(x) = (x+ 0.4 x^2) 1_{x>0}$}
 \end{minipage}
\begin{minipage}[b]{0.5\linewidth}
  \centering
 \includegraphics[width=0.9\textwidth]{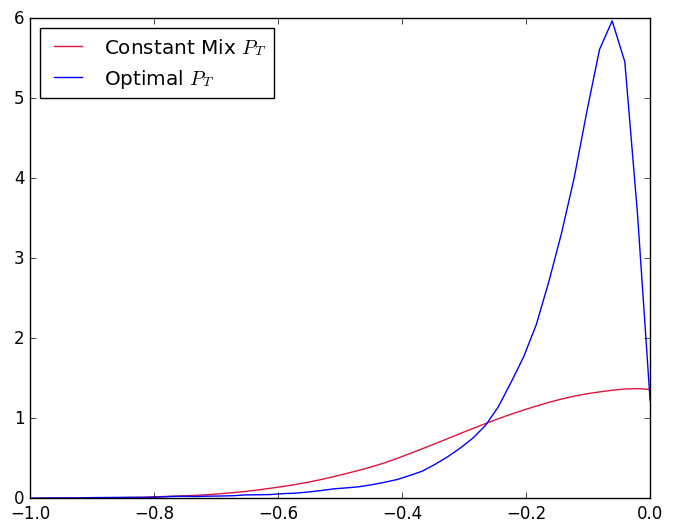}
 \caption*{$g(x) = (x+ 0.4 x^2+ 4e^{-4} x^3) 1_{x>0}$}
 \end{minipage}
\begin{minipage}[b]{\linewidth}
  \centering
 \includegraphics[width= 0.45\textwidth]{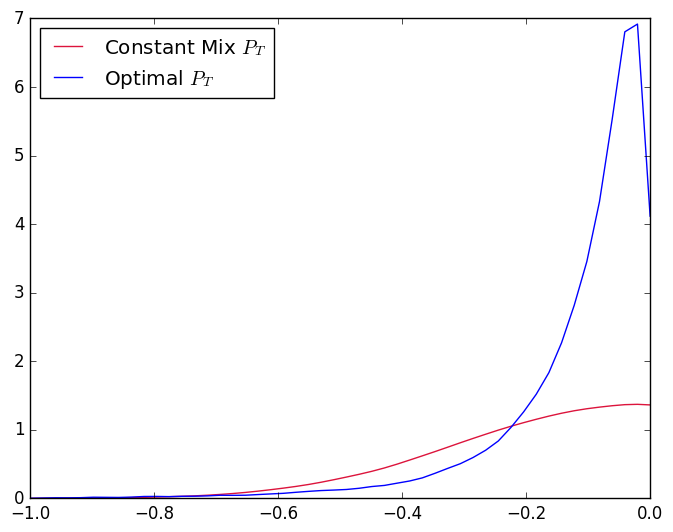}
 \caption*{$g_3(x) =(x+ 0.4 x^2+ 4e^{-5} x^3) 1_{x>0}$}
 \end{minipage}
\caption{Normalized $P_T$ obtained for different objective functions with the Black Scholes model;} 
\label{figFuncOptimBS}
\end{figure}

For these three normalized functions we give the different quantiles obtained in table \ref{quatOptBS}.

\begin{table}[!h]
 \centering
 \begin{tabular}{|c|c|c|c|c|}  \hline
Quantile & \multicolumn{3}{|c|}{Optimization} & Constant Mix \\ \hline
Objective         & $g_1(x)$  & $g_2(x)$& $g_3(x)$&  \\ \hline
1\% &    -0.72380 &  -0.57396  &  -0.62012    &    -0.62889\\ \hline
2\% & -0.59356   & -0.477630 & -0.51152       &  -0.55947     \\ \hline
5\% &  -0.37616  &  -0.35093 &  -0.35795 &  -0.452577  \\ \hline
10\% &  -0.21702 & -0.26972  &  -0.25986 &  -0.354676  \\ \hline
20\% & -0.11440 &   -0.19472 & -0.17181  &   -0.236149 \\ \hline
30\% &  -0.07264&  -0.1533   & -0.12504  &   -0.14848 \\ \hline
\end{tabular}
\caption{ \label{quatOptBS} Normalized $P_T$  quantile obtained with Black Scholes model with different objective functions.}
\end{table}
As expected, it seems to be impossible be get a loss distribution always better than the one obtained Constant Mix. Penalizing the tail as done with function $g_2$ permits to get far better results for extreme quantile but degrades the distribution much of the time. With the $g_1$ function the optimal distribution is above the Constant Mix in $42\%$ of the case, while with the $g_2$ function it is  around $30 \%$, and with the $g_3$ function it is $35\%$. With $g_2$ and $g_3$, the optimal solution renounces to gains and prefers to secure some quite small losses.
In the sequel we restrict ourself to the $g_3$ function which seems to fit the practical risk aversion.

\subsection*{Towards an heuristic equivalent to the optimal strategy}
\begin{figure}[!h]
\begin{minipage}[b]{0.45\linewidth}
  \centering
 \includegraphics[width=0.9\textwidth]{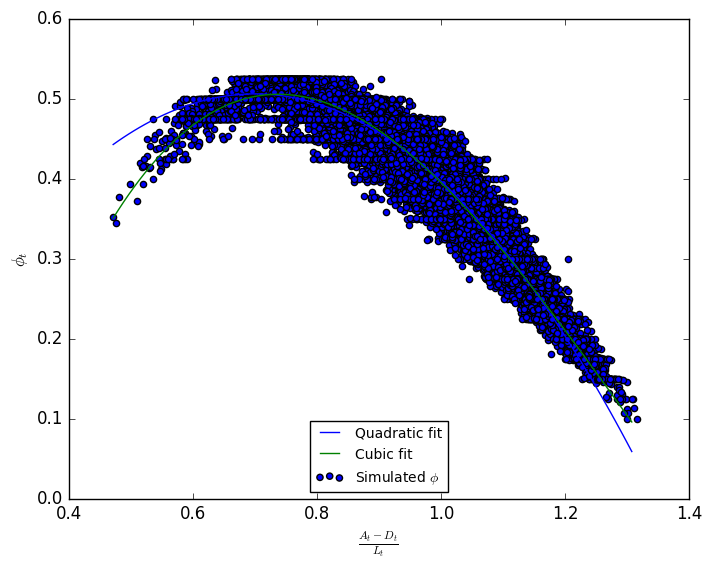}
 \caption*{$\phi_t$ with $t=3$ years}
 \end{minipage}
\begin{minipage}[b]{0.45\linewidth}
  \centering
 \includegraphics[width=0.9\textwidth]{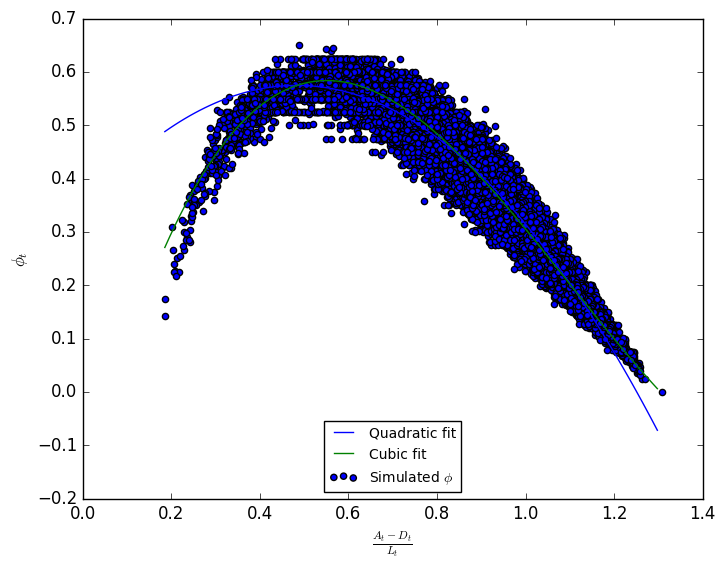}
 \caption*{$\phi_t$ with $t=8$ years}
 \end{minipage}
  \centering
 \includegraphics[width=0.45\textwidth]{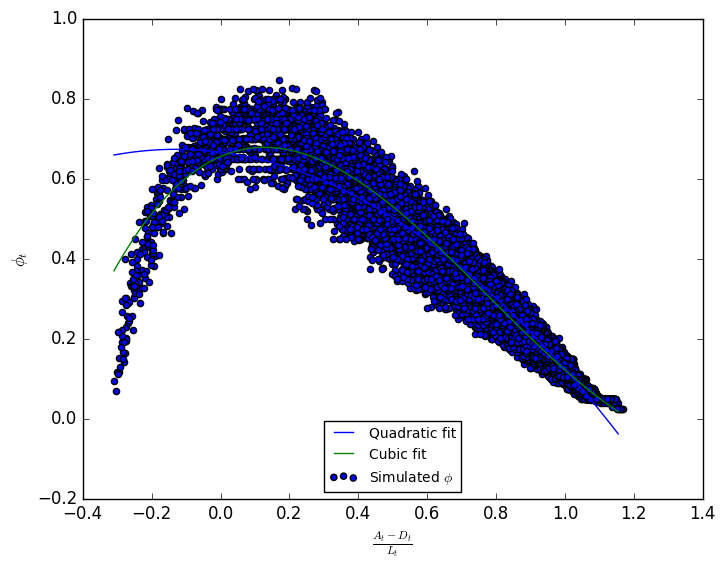}
 \caption*{$\phi_t$ with $t=15$ years}
\caption{$\phi_t$ simulation for different dates $t$.} 
\label{figFitBS}
\end{figure}

\begin{figure}[!h]
\centering
\includegraphics[width=0.9\textwidth]{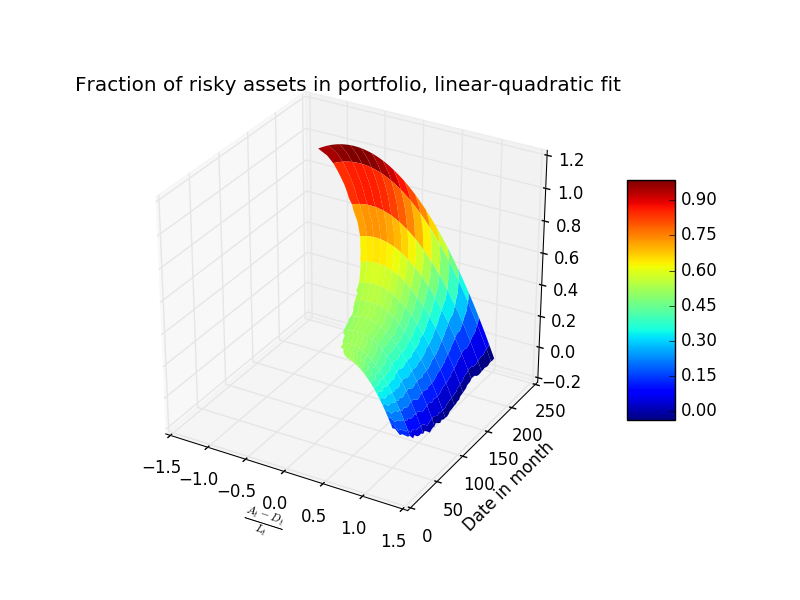}
\caption{\label{fighatphiApprox}Linear-Quadratic  fit of the strategy, Black Scholes model}
\end{figure}

\begin{figure}[!h]
\centering
\includegraphics[width=0.7\textwidth]{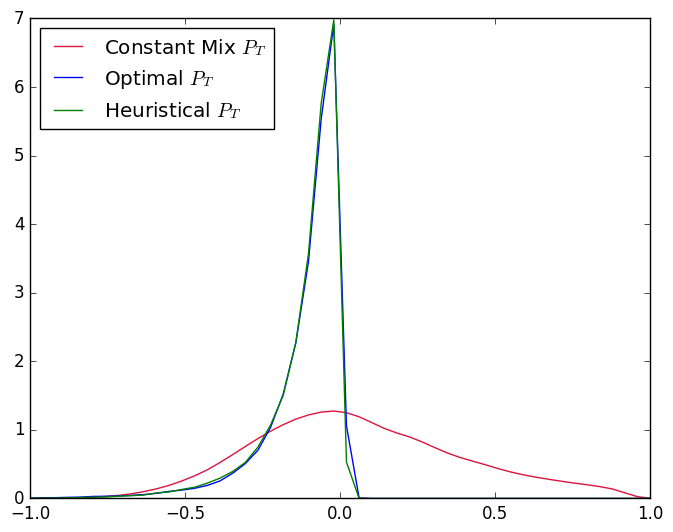}
\caption{\label{fighatphiApprox1} Normalized $P_T$ distribution with Constant Mix, optimal and Linear-Quadratic heuristic strategy with the Black Scholes model.}
\end{figure}

Practitioners need to understand the strategy associated to an optimization problem in order to trust it and to check its robustness.
We focus for example on the strategy associated to the $g_3$ function and we try to explicit the strategy as a function of the funding ratio $\frac{A_t-D_t}{L_t}$
so we approximate $\phi_t$ the proportion of the portfolio invested in the equity index by :
\begin{equation}
 \phi_t \simeq F \left (t, \frac{A_t-D_t}{L_t} \right ).
\label{approxPhiBS}
\end{equation}
On figure \ref{figFitBS} we give  the  strategies at different dates and show that the strategy can be accurately described by a quadratic or cubic fit at each date using some linear regressions.
The optimal strategy at each date consists in de-risking the portfolio when the funding ratio is satisfactory  and taking more risk
when the funding ratio degrades. This optimal strategy can be linked to the de-risking strategies observed with concave strategies.
Notice that when the funding ratio becomes very bad and as the time to maturity gets smaller, the strategy consists in 
de-risking  again the portfolio showing that there is no hope to recover the losses.
Note that for very high funding ratio the whole portfolio is invested in bonds.
According to the preceding results, we look for a linear in time quadratic in space approximation of the strategy :
\begin{equation}
F(t,x) = (a_0+a_1 t) +  (b_0+b_1 t) x + (c_0+c_1 t) x^2.
\label{hatphiApprox}
\end{equation} 
Figure \ref{fighatphiApprox} gives the resulting $F$ function and figure \ref{fighatphiApprox1} shows that the Linear-Quadratic heuristic is very effective. 

\begin{figure}[!h]
\centering
    \includegraphics[width=0.6\textwidth]{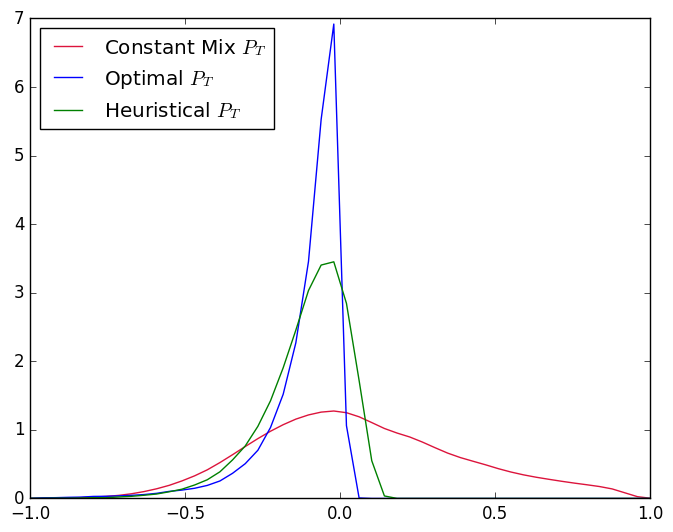}
\caption{\label{fighatphiQuadSimpleApprox} Comparison of Constant Mix and simplified heuristic  normalized $P_T$ distribution}
\end{figure}

At last we can observe on figure \ref{figFitBS} that the optimal strategy is not very dependent on time so we can search for a simple quadratic
approximation of the strategy using linear regression. The fitted solution is given by :
$$F(t,x) :=F(x)= -0.113 x^2  -0.377 x + 0.731.$$
Using this simplified heuristic, we give the $P_T$ distribution associated to this strategy on figure \ref{fighatphiQuadSimpleApprox}.
The de-risking strategy obtained allows us to gain in the tail as shown on table  \ref{fighatphiQuadSimpleApprox1} while not renouncing to some gains : the resulting distribution is more satisfactory from the practitioner point of view (see for example the $1\%$ quantile )
than the ``optimal'' one  indicating  that the objective function could be improved.

\begin{table}[!h]
\centering
\begin{tabular}{|c|c|c|}  \hline
Quantile & Simplified  & Constant  \\ 
 &  heuristic &  Mix \\ \hline
1 \%&  -0.55169  & -0.62889\\ \hline
2\% & -0.477519  &  -0.55947     \\ \hline
5\% & -0.37311 &  -0.452577  \\ \hline
10\% &  -0.29359 &  -0.354676  \\ \hline
20\% &  -0.20869 &   -0.236149 \\ \hline
30\% & -0.15636 &   -0.14848 \\ \hline
\end{tabular}
\caption{\label{fighatphiQuadSimpleApprox1}Comparison of Constant Mix and simplified heuristic normalized $P_T$  quantiles.}
\end{table}

\subsection*{Robustness of the strategy}
\begin{figure}[!h]
\centering
\includegraphics[width=0.6\textwidth]{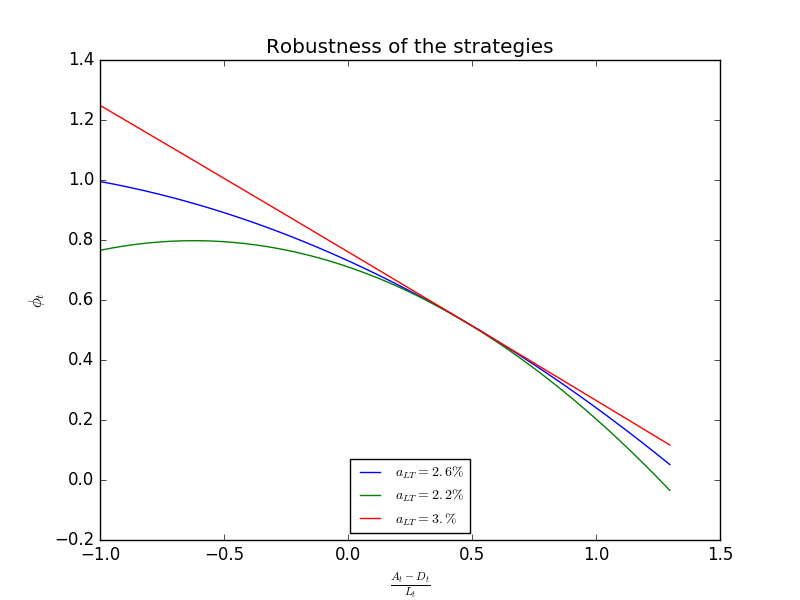}
\caption{\label{figRobustStrag}Testing the  sensibility  to  $a_L$ on the strategy.}
\end{figure}

\begin{figure}[h!]
\centering
\includegraphics[width=0.6\textwidth]{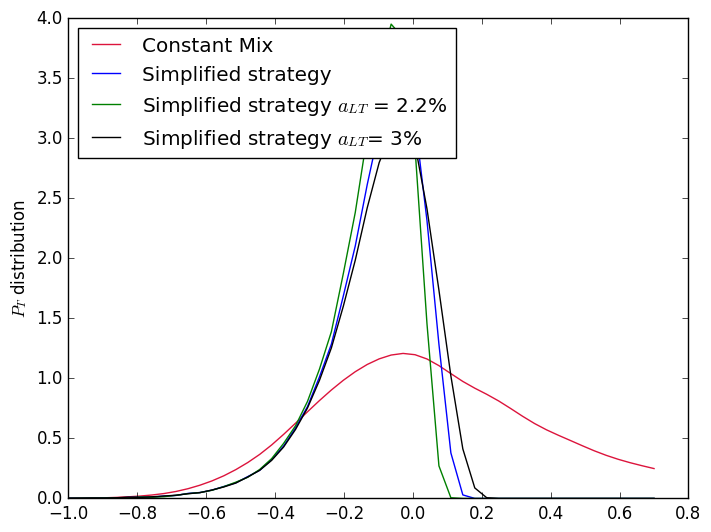}
\caption{\label{figRobustStrag1}Testing the  sensibility  to  $a_L$ on the $P_T$ distribution.}
\end{figure}

As asserted in the introduction, the principal risk in the short term comes from the long term actualization factor $a_{L}$ that has to be actualized due to the constraint to be below the TEC 30 average value on the last 48 months.
We test the robustness of the simplified strategy to a change in the $a_L$ value. We still suppose that the 'real' $a_L$ value  is equal to $2.6\%$ but that the optimization has been achieved with some different $a_L$ values equal to $2.2\%$ or $3\%$. The simplified constant in time quadratic strategy is identified for the two values and reported on figure \ref{figRobustStrag}  : as expected the two other strategies are de-risking strategies. 
On figure \ref{figRobustStrag1}, we plot the normalized  $P_T$ distributions using the three simplified heuristics and an  $a_L$ value  equal to $2.6\%$.
The figure  \ref{figRobustStrag1} indicates that the de-risking strategy is stable with respect to the $a_L$ parameter. 

 
\newpage
\section{Optimal strategy for an index following the MMM model}
In this section we take a second modelling for the dynamic of the asset. This modelling is more adapted to long term valorization and will permit to check the robustness of the strategy previously obtained with respect to the model.
\subsubsection*{The MMM model}
\begin{figure}
\centering
\begin{subfigure}[b]{0.45\textwidth}
  \centering
 \includegraphics[width=\textwidth]{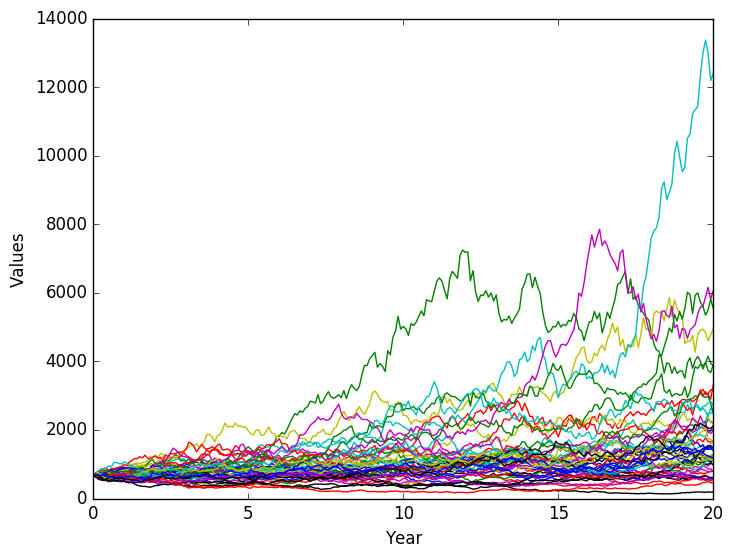}
\caption{Some Black Scholes  simulations}
\end{subfigure}
\begin{subfigure}[b]{0.45\textwidth}
  \centering
 \includegraphics[width=\textwidth]{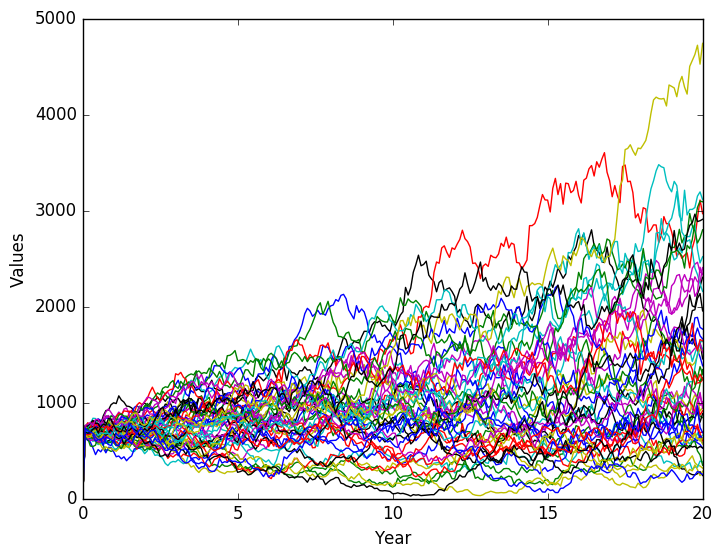}
\caption{Some MMM simulations}
\end{subfigure}
\caption{\label{compBS_MMM} Some simulations using  the Black Scholes and MMM model.}
\end{figure}

This section will allow us  to test the strategy  robustness with respect to the model used for the equity index.\\ 
The Black Scholes model  suffers from a lot of shortcomings :
\begin{itemize}
\item many studies seem to indicate that the volatility is stochastic with a volatility correlated to the asset \cite{Platen} explaining the development of 
models with stochastic volatility \cite{Heston}, \cite{Hull}, \cite{Stein},
\item the observed log returns are not normal giving a  skewed distribution with heavy tails explaining the use of jumps and Levy processes to model the prices  \cite{Kou} \cite{Merton3} \cite{Cont}.
\end{itemize}
But most of all, the long term implicit volatility is far too high yielding for example very high option prices.
The success of these models and in particular of the Black Scholes model  is due to their  Absence Of Arbitrage (AOA) property,  equivalent  to the
existence of martingale measure allowing to give a price to financial assets.
As explained in \cite{Platen}, the AOA can be relaxed still giving useful models with arbitrage impossible for strategies keeping positive portfolio values.
A similar framework has been developed in \cite{Fernholz} and an unified view of these approaches can be found in \cite{Fontana}.\\
In \cite{Platen}, the Minimal Market Model (MMM) has been proposed to model the dynamic of the actualized Growth Optimal Portfolio (GOP). 
Its dynamic is given by :
\begin{equation}
d  S_t    =    \alpha_t dt + \sqrt{  S_t \alpha_t} dW_t,
\end{equation}
with $\alpha_t = \alpha_0 e^{\eta t},$ where the exponential form permits to model the exponential growth in the economy.\\
It is showed in \cite{Platen}
that an equi-weighted portfolio with a large number of assets is a good candidate to approximate the GOP. 
In \cite{Platen}, it is shown for example that the MMM correctly fits the WSI.
In the sequel we suppose that our risky index is the  MSCI World and that the MMM is a good candidate to modelize the MSCI world actualized dynamic.
\begin{eqnarray*}
d  S_t  & = &  \alpha_t dt + \sqrt{  S_t \alpha_t} dW_t. 
\label{eq:MMM}
\end{eqnarray*}
The advantages of this model are the following:
\begin{itemize}
\item it is parsimonious, with only two parameters to fit,
\item due to this parsimony, we have a stable and easy way to calibrate the parameters  using the empirical covariance of $\sqrt{S_t}$ \cite{Platen},
\item the model can be simulated exactly - noting that the equation \eqref{eq:MMM} is a time transformed squared Bessel process of dimension 4 \cite{Platen} - using the procedure described in  \cite{Platen3}, \cite{Platen4}, \cite{Platen5},
\item It keeps the properties of the stochastic volatility models giving high volatilities when the prices is low and low volatilities when the prices are high.
\end{itemize}
The MMM model has been calibrated on the MSCI World between  1976-2016 yielding the parameters :
 \begin{eqnarray*}
\alpha_0  & \simeq &   2.317 \nonumber  \\
\eta      & \simeq &  0.0542  \nonumber \\
\end{eqnarray*}
On figure \ref{compBS_MMM} we simulate some prices obtained with the MMM model and the Black Scholes model with $50$ paths  showing that the Black Scholes
 model gives very high values with a high probability.
\subsubsection*{Distribution with Constant Mix Strategy}
\begin{figure}[!h]
\centering
\includegraphics[width=0.6\textwidth]{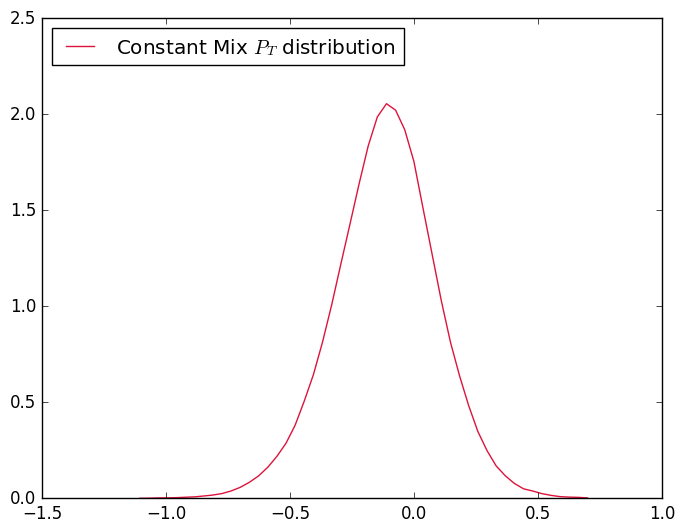}%
\caption{\label{CMixDistMMM} Normalized $P_T$ distribution of the Constant Mix strategy for the MMM (using Black Scholes renormalization factor).}
\end{figure}

The distribution obtained by the Constant Mix  strategy for the Asset Liability problem  is  given on table \ref{CMixDistMMM}.

\begin{table}[!h]
\centering
\begin{tabular}{|l|l|l|}\hline
                  & BS & MMM \\ \hline
    Quantile 1\%  &  -0.62889 &  -0.64045 \\ \hline
    Quantile 2\%  &  -0.55947 &  -0.56887 \\ \hline
    Quantile 3\%  &   -0.51350 & -0.52499 \\ \hline
    Quantile 5\%  &  -0.45257 &  -0.46535 \\ \hline
    Quantile 10\% &  -0.35467 &  -0.37876 \\ \hline
    Quantile 20\%  & -0.23614 &  -0.28117  \\ \hline
  \end{tabular}
  \caption{ \label{CMixQuantMMM} Normalized $P_T$ quantiles of the Constant Mix strategy with the Black Scholes  and the   MMM  (using Black Scholes renormalization factor).}
\end{table}

\subsubsection*{Objective function}
Comparing the quantiles  obtained  in table \ref{CMixQuantMMM} for the MMM and the   Black Scholes model  we notice that the quantiles for the losses
up to $10\%$ are very close but as expected that  the distributions of the gains are very different : due to the volatility structure of the model, a lot of Black Scholes
simulations give very unrealistic high returns.\\
With the MMM dynamic, the portfolio follows the following Stochastic Differential Equation :
\begin{eqnarray}
d A_t =  \frac{A_t}{S_t}  (  \alpha_t dt + \sqrt{  S_t \alpha_t} dW_t) +  dD_t,
\end{eqnarray}
thus  the objective function given by equation \eqref{eq:objective} is a function of $t$,$A_t$,$S_t$ and $D_t$, and the optimization problem becomes :
\begin{equation}
\hat J(t, A_t, S_t, D_t) =  \min_{\phi} \esp{ g(-P_T) ~|~\Fc_t}.
\label{Eq:OptimMMM}
\end{equation}
\subsubsection*{Numerical results}
\begin{figure}[!ht]
\centering
\includegraphics[width=0.8\textwidth]{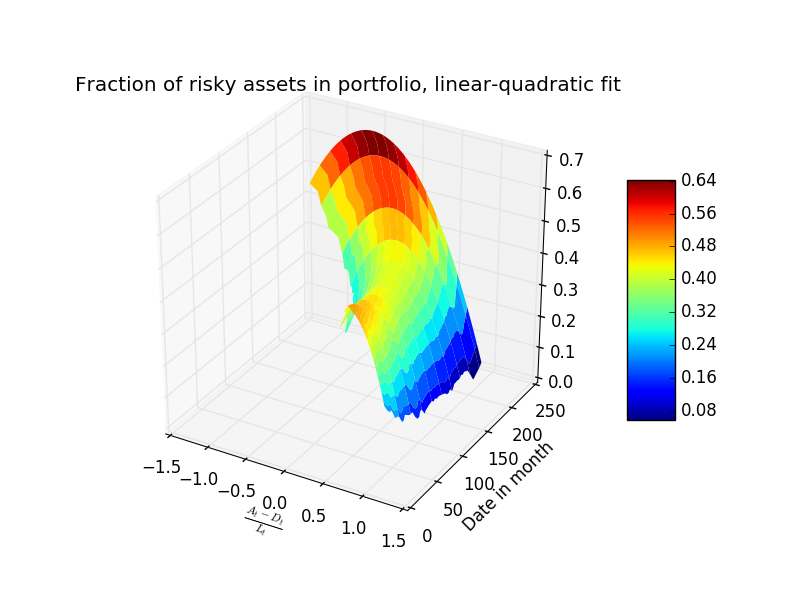}
\caption { \label{MMMLinearQuad} Linear-Quadratic fit of the optimal strategy with the MMM.}
\end{figure}

\begin{figure}[!ht]
\centering
\includegraphics[width=0.6\textwidth]{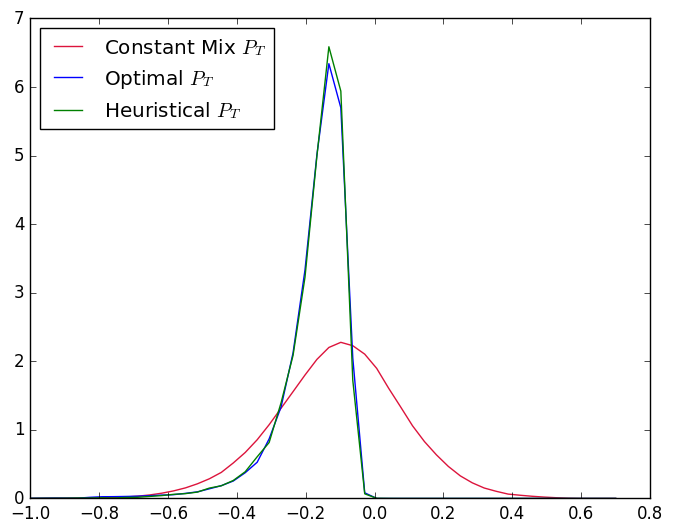}
\caption {\label{MMMLinearQuadDistrib} $P_T$ distribution with Constant Mix, optimal and Linear-Quadratic heuristic strategy with the MMM.}
\end{figure}

The numerical resolution of \eqref{Eq:OptimMMM} compared to \eqref{Eq:OptimBS}  involves a four dimensional problem  much harder to tackle with Monte Carlo methods. The risky asset values
 $S_t$  are  discretized  with $100$ meshes defined by the partition of the values created with  $600 000$  samples as explained in \cite{StOpt}.
All distributions are calculated by simulating the strategies with $50 000$ samples.\\
We keep using $g_3$ for the $g$ function in the objective function \eqref{Eq:OptimMMM}.
Once again we fit the optimal strategy as a function of the funding ratio according to \eqref{approxPhiBS} using a linear in time quadratic in space approximation.
The $F$ function obtained is given by figure \ref{MMMLinearQuad}.  Using this Linear Quadratic heuristic strategy, the $P_T$ distribution is given on figure  \ref{MMMLinearQuadDistrib} and
is very similar to the one obtained by optimization.


The strategies obtained (optimal and Linear-Quadratic)  are very similar to the ones obtained with the Black Scholes model : they are   de-risking strategies.
 They de-risk the portfolio if the funding ratio is good and take some risk if the funding ratio is bad.\\

\begin{figure}[h!]
\centering
\includegraphics[scale=0.4]{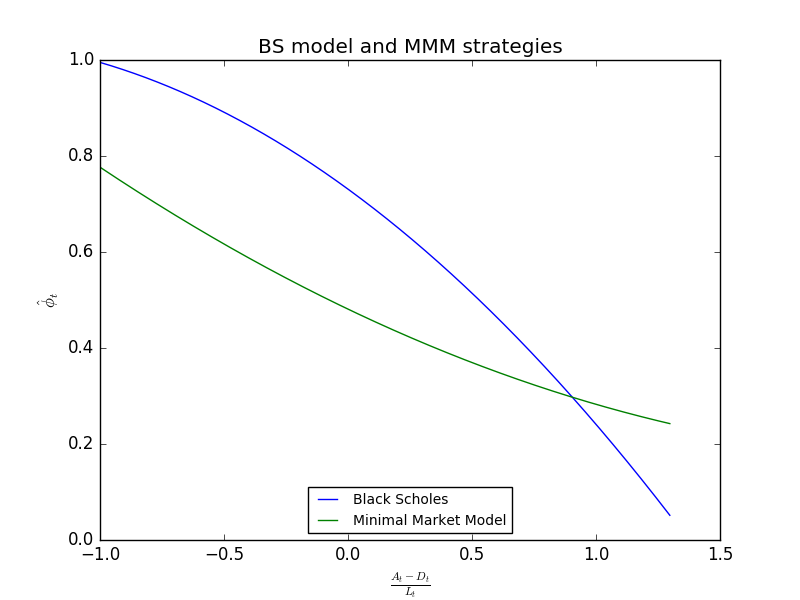}
\caption{ \label{QuadStratBSMMM} Comparison of the two Quadratic strategies obtained by the MMM and the Black Scholes model. }
\end{figure}

Once again a quadratic independent in time strategy can be computed by regression. A comparison of the de-risking heuristic quadratic strategies
obtained by Black Scholes model and the MMM can be found on figure \ref{QuadStratBSMMM}. At fist order, the strategies are the same.
Once again the quadratic strategy give good results for the tails as shown on  table \ref{MMMQuantile}.

\begin{table}[h!]
\centering
\begin{tabular}{|c|c|c|c|c|} \hline
Quantile &   Constant Mix & Optimized &  Linear-Quadratic &  Quadratic  \\
         &            &              & heuristic         &   heuristic         \\ \hline
1\% &   -0.64045   &   -0.631190 &   -0.57322 & -0.52199\\ \hline
2\% &   -0.56887   &   -0.517330 &  -0.495545 &   -0.469857 \\ \hline
5\% &   -0.52499   &   -0.394088 &   -0.38883 & -0.394860 \\ \hline
10\% &  -0.46535   &  -0.31676 &  -0.31524 &  -0.3374 \\ \hline
20\% &  -0.37876   &  -0.25002 &   -0.24987 &  -0.2779036\\  \hline
30\% &  -0.28117   &  -0.21365 &  -0.21256 & -0.23951\\  \hline
\end{tabular}
\caption{\label{MMMQuantile} $P_T$ quantiles by optimization, Constant Mix, Linear-Quadratic Heuristic and Quadratic Heuristic in millions of Euros.}
\end{table}

\begin{figure}[!ht]
\centering
\includegraphics[scale=0.5]{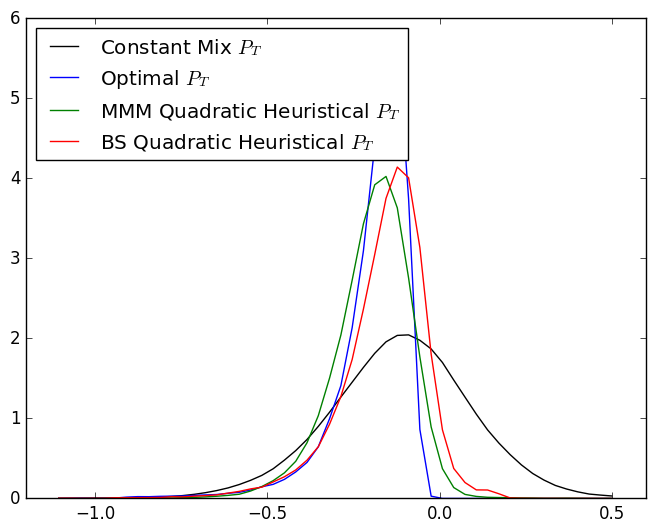}

\caption{ \label{QuadDistrMMM} Comparison of distributions obtained  on the MMM by the Constant Mix strategy, the optimal strategy, the Quadratic heuristic strategy derived from the MMM optimization and Quadratic strategy calibrated on the Black Scholes model.}
\end{figure}

At last the figure \ref{QuadDistrMMM} allows  to compare the distribution obtained with the MMM model  using the Constant Mix strategy, the one obtained by optimization,
the Quadratic strategy calculated, and the Quadratic strategy calculated using a Black Scholes model.
It shows that the result are quite robust with respect to a structural change in the model.

\section{Conclusion}
We have found that the ALM problem of the French Nuclear Operator can be cast as an optimization problem where the function to optimize penalizes the endowment realized
 in order to respect the regulatory constraint on the liability. We have shown that the strategy obtained can be easily explained by a de-risking behaviour which
is robust with respect to the long term discount factor  affecting the liability and that this strategy is also robust to the model choice.

\end{document}